# Detection and characterization of colloidal silver nanofluids by photothermal techniques

M. S. Swapna,[1,2,*] S. Sankararaman,[2] and D. Korte[1,3]

[1]*Laboratory of Environmental Research, University of Nova Gorica, Vipavska 13, 5000 Nova Gorica, Slovenia*
[2]*Department of Optoelectronics, University of Kerala, Trivandrum 695581, Kerala, India*
[3]*Dorota.Korte@ung.si*
*\*swapnams.opto@gmail.com*

**Abstract:** In this work $Ag^0$ nanoparticles (NPs) were synthesized and detected by flow injection analysis coupled to collinear dual beam thermal lens spectrometric (TLS) detection. The estimated limit of detection was 0.8 µg/L. The use of the IonPac Cryptand G1 column enabled $Ag^0$ NPs detection in the presence of interfering ions normally presented in water. $Ag^0$ nanofluids (NFs) were further characterized by time resolved TLS and beam deflection spectrometry to determine the NFs thermal diffusivity and conductivity. The applied methods were found to be fast, simple, reliable, and highly sensitive.

## 1. Introduction

Silver nanofluids ($Ag^0$ NFs) are stable colloidal suspensions obtained by dispersing of ionic silver's particles', its compounds or silver colloids with particles of a size below 100 nm in base fluid. This feature is the result of Brownian motion that the dispersed NPs undergo and consequently frequently collide with the heat source leading to heat exchange between the NPs and heat source that occurs very rapidly within the time of few ps. Thus, the temperature of NPs is nearly instantaneously raised and dropped as it is enhanced by the NP's small size as well as high surface to volume ratio what further increases the heat conduction in the NF since its transfer is a surface phenomenon. Unfortunately, the added solid NPs into the base fluid tend to precipitate. A solution for this problem is the use of colloidal solutions to keep the NF stable with NPs uniformly distributed within it and causing neither clogging nor erosion. Because of that, the nanofluids found application in many fields of industrial technology (e.g. microelectronics, transportation, chemical production, heat exchangers, cooling equipment) [1]. Recently, the colloidal silver ($Ag^0$) NFs were found to be promising material in application where enhanced heat transfer performance is required [2]. Furthermore, the NF function is strongly dependent on the NP volume fraction [3,4]. The higher NPs concentrations is, the better heat transfer performance is obtained [5] but this effect is valid only beyond a certain NPs concentration over which the heat transfer start to deteriorate as a result of NPs aggregation and/or sedimentation. Thus, it is of high importance to evaluate the way how NPs concentration affects the functionality of NF in thermal engineering applications, thus, maximize its heat transfer potential [6].

Colloidal silver NFs not only have enhanced thermo-physical properties but also were found to be antiseptic, antibacterial and antimicrobial what makes them be used as a disinfectant for the drinking water supplies in developing countries and the International Space Station (ISS) [7]. $Ag^0$'s disinfection mechanism includes inactivation of bacteria and mold cell enzymes that need oxygen for their metabolism; it causes their cellular disruption. It was found recently that silver based NPs are capable of deactivating up to 99%–100% of bacteria and viruses from drinking water [8]. Colloidal silver can remain in water for a long period of time, but is not considered to have good residual power because of the slowness of its reactions in eliminating organic matter. Furthermore, in large amounts it is toxic and its ingestion leads to blue-gray discoloring of skin (argyria), deposition of silver in eye (agryrosis),  serious neurologic, renal, or hepatic complications and even death [9]. It may also reduce the absorption of some medications such as tetracycline, quinolone antibiotics and penicillamine reducing their effectiveness. The recommended dose for high germicidal efficacy is in the range of 25 – 75 µg/L whereas the maximum contamination level (MCL) is 50-100 µg/L for water disinfection in different countries and 200-500 µg/L in case of space missions [10]. The MCL of silver is defined by the secondary drinking water standards which are referred to those parameters that may impart an objectionable appearance, odor or taste to water, but are not health hazards when present in water in excess of the MCL as it is in case of the primary drinking water standards [11].The technology of applying the colloidal silver as disinfection agent has been developing for decades by the National Aeronautic and Space Administration (NASA), European Space Agency (ESA) [12] and World Health Organization but for effective utilization of such technology in practice, a sensitive water quality monitoring system is needed.

Recently, photothermal techniques (PT) were found to be a promising alternative for determination and characterization of a wide range of species due to their higher sensitivity, reliability and effectivity [13–15]. Thus, in this work characterization of $Ag^0$ NFs is performed by PTs. The concentration of colloidal silver in NFs is investigated by the use of thermal lens spectrometry (TLS) whereas the effect of $Ag^0$ concentration on the NF's thermal properties is studied by beam deflection spectrometry (BDS)[16].

TLS provides more sensitive detection of nanosilver compared to conventional absorbance measurements and more convenience than Surface-enhanced Raman scattering (SERS) [17,18], electrochemical detection (ED) [19] or femtosecond laser spectroscopy (FLS) [20]. TLS coupled to flow injection analysis (FIA) was applied for direct detection of nanosilver and other silver species in the water supplies [21]. The FIA was already applied to the colloid silver production in a flowing $AgNO_3$ solution [22] where $Ag^0$ was formed by photoreduction or by $NaBH_4$ in its flowing solution [17] what eliminates the need of stabilizers or polymers which can interfere with the metal surface and influence the results of analysis [17].

BDS enable the simultaneous determination of optical, thermal properties and related properties of solids, liquids, gases, thin films, multilayered structure, both composite and colloidal materials in a non-contact and non-destructive way with a high signal-to-noise ratio [23–25]. It provides the ability of material analysis in the amount of nanograms what is a key important requirement for a modern, high precision sensing technology [26–28]. Thus, BDS technique becomes suitable for studying thermal properties of precious analytes including colloidal NFs with foreseen application in microelectronic industry [29].

The aim of this study is to apply the FIA with TLS detection [30] to monitor the amount of $Ag^0$ in NF as well as time resolved TLS to determine the NF thermal diffusivity and BDS to find its thermal conductivity as a function of $Ag^0$ concentration. Furthermore, investigation of processes occurring during the analysis due to interaction of the silver ions with ions normally presented in water, is also carried out in order to reduce the contribution of interfering foreign ions on the results. In such a way the effect of $Ag^0$ concentration in NF on its heat transfer performance is presented what allows us to determine the $Ag^0$ NF optimum values of thermal properties for its application in different thermal systems.

## 2. Materials

### 2.1 Reagents and solutions

The following chemical were used as purchased without further purification: $AgNO_3$, NaCl, and $CuSO_4·5H_2O$, $NaBH_4$, NaOH were from Carlo Erba, $KNO_3$, $Pb(NO_3)_2$, and $3CdSO_4·8H_2O$ were from Riedel DeHaen, $MgSO_4$, $Na_2CO_3$, and $MnSO_4·H_2O$ were from Fluka, NaI from Fluka, NaBr, $NH_4OH$ 25% from J.T. Baker, $Na_2SO_4$ from Alfa Aesar, $Na_3PO_4$ from Scharlau, $Fe_2(SO_4)_3·3H_2O$ from Riedel De Haen, $ZnCl_2$ from Laphema, $Hg(NO_3)_2·H_2O$ from Kemika and $HNO_3$ 65% from MERCK KGaA. All solutions were prepared fresh daily using high purity double-deionized H$_2$O (18 MΩ/cm) obtained with Purelab Option-Q and protected from light by covering them with an aluminum foil.

*2.2 Silver colloids formation*

The generation of silver colloids is based on a chemical reduction of silver ions by borohydride that follows the reaction:

$$2AgNO_3 + 2NaBH_4 + 6H_2O \rightarrow 2Ag + 7H_2 + 2NaNO_3 + 2H_3BO_3 \qquad (1)$$

For constructing the calibration curve of the experimental setup, the stock solution of Ag$^+$ (1 mg/mL) was prepared by dissolving 15.74 mg of AgNO$_3$ in 10 mL water, on its basis the working solutions with silver ions concentrations in the range of 20 – 2500 µg/L were prepared by appropriate dilutions of the stock solution in 10 mL water. As a reductant the 0.6 mM sodium borohydride at pH 12.5 was used. For that purpose, a water solution of NaOH was prepared by dissolving 5 g of NaOH in 95 mL of water, then 8.2 mL of such solution was mixed with 500 mL of water to reach water solution of NaOH at pH 12.5. To obtain 0.6 mM concentration of the reductant, the stock solution of 0.1 M sodium borohydride at pH 12.5 was prepared by dissolving 0.095 g of NaBH$_4$ in 25 mL of aqueous solution of NaOH at pH 12.5 and diluting 3 mL of obtained solution further into 500 mL of NaOH solution at pH 12.5. As a result of presence of van der Waals attraction between Ag$^0$ NPs, they tend to aggregate forming bulk particles. To prevent that process, the synthesized NPs are stabilized by protective layers that in our study is the layer of adsorbed borohydride ions (Fig. 1).

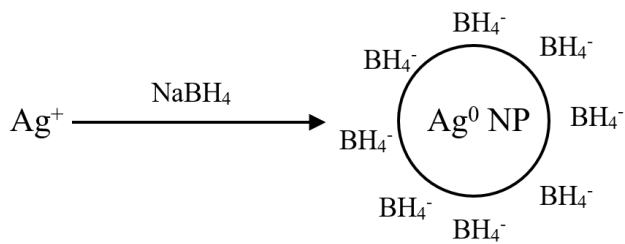

Fig. 1. Ag$^0$ NPs formation and creation of adsorbed borohydride ions layer around them.

It must be noticed that NaBH$_4$ works both as a reductant of Ag$^+$ to Ag$^0$ and as a stabilizer of Ag$^0$ by adsorbing on its surface. The NaBH$_4$ concentration should be at least twice higher comparing to that one of AgNO$_3$. If this condition is not fulfilled the aggregation of Ag$^0$ NPs occurs as a result of increased ionic strength.

*2.3 Chemicals for investigation of interfering ions*

The stock solutions of foreign ions were prepared by dissolving appropriate amounts of corresponding salts in 10 mL water. The working solutions with 1, 10 and 100 mg/L ions concentrations were prepared by dilution of 10, 100 and 1000 µL stock solutions in 10 mL of double deionized water and adding to such a solution 100 µL (in case of TLS detection) or 1 mL (for spectrophotometry absorption measurements) of Ag$^+$ stock solution. Additionally, in spectrophotometry experiments, 1 mL of reducing agent was added to each of the working solutions.

*2.4 Reagents for IonPac Cryptand G1 column testing*

The working solutions with 1000, 250 and 100 µg/L of Ag$^+$ concentration was prepared by dilution of 100 µL of its stock solutions in 10 mL of double deionized water to receive solution of 10 mg/L, further 1000, 250 or 100 µL of such a solution was diluted in 10 mL of H$_2$O to get the working solution of 1000, 250 and 100 µg/L of Ag$^+$ concentration, respectively. The working solutions with 300 µg/L Fe$^{3+}$, 50 µg/L Mn$^{2+}$ and 1.3 mg/L Cu$^{2+}$ concentration were prepared by dissolving the 100 µL of their stock solution in 10 mL of double deionized water and 300 µL (in case of $Fe^{3+}$), 50 µL ($Mn^{2+}$) and 1.3 mL ($Cu^{2+}$) of such a solution were further dissolved in 10 mL of $H_2O$ to get the desired concentration to which 100 µL of 10 mg/L $Ag^+$ concentration was added. The $NH_4^+$ solution at pH 11 was prepared by adding dropwise concentrated ammonium hydroxide (25% NH4OH) to double deionized water and controlling the pH by pH-meter (HI 8417, HANNA instruments) constantly mixing using magnetic stirring (LMS-1003, IDL GmbH&Co. KG).

## 3. Method description

### 3.1 Thermal lens spectrometry

In a sample that is periodically heated by a modulated laser light beam (excitation beam EB) due to the nonradiative loss of absorbed energy, a temperature gradient and a consequent refractive index gradient (thermal lens) is generated which can be detected by another laser beam (probe beam PB). The thermal lens changes the light intensity distribution in PB cross-section which are directly related to the absorption $A$ of the sample. For the sample located in the experimentally determined optimal position for achieving the maximum TLS signal, which is approximately $3^{1/2}$ of the confocal distance $Z_c$ beyond the probe beam waist, the relative changes of the probe beam intensity resulting from its interaction with the thermal lens can be described as [31]:

$$\Delta I/I = 2.303EA + (2.303EA)^2/2 + \ldots \qquad (2)$$

where the enhancement relative to the transmission measurements E is given by:

$$E = -P(dn/dT)/1.91\lambda k \qquad (3)$$

and $\lambda$ the probe beam wavelength and $P$ the excitation laser power, while $Z_c = 2\pi a^2/\lambda$ is the confocal distance, $k$ is the thermal conductivity of the solvent and $a$ the probe beam radius.

For low-concentration or weak absorbing species resulting in $A < 0.1$ the second term in eq. (2) can be neglected. The TLS signal is then directly proportional to the absorbance, excitation power and enhancement factor of the sample what enables this technique to be used as a highly sensitive method for the determination of chemical species at very low concentration.

### 3.2 Flow-injection analysis (FIA) system with thermal lens spectrometric (TLS) detection unit

To determine the presence of ionic silver in water by chemical reduction with $NaBH_4$ a home-built FIA system with a dual-beam TLS detection unit was used. The instrumental set-up is schematically presented on Fig. 2.

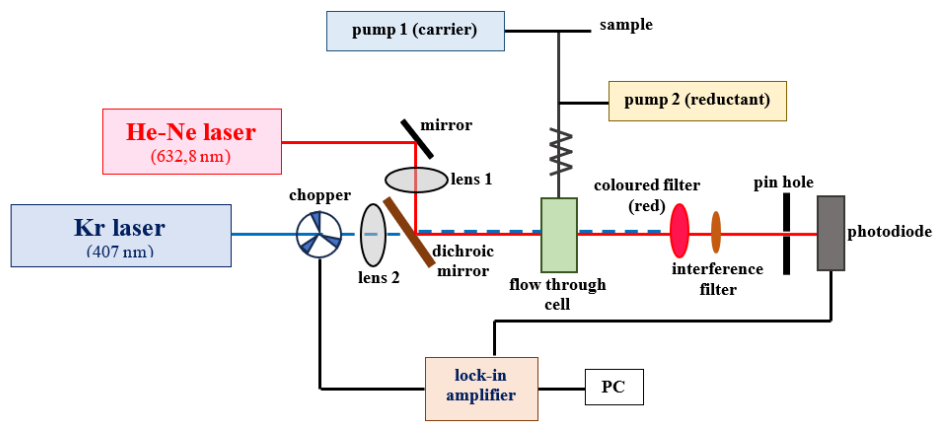

Fig. 2. FIA-TLS experimental setup for NFs $Ag^0$ concentration determination.

The FIA system consists of two HPLC pumps: one of them (Shimadzu, model LC10Ai), used to deliver the reducing reagent and the other (KNAUER), used to pump the carrier solution transporting the $Ag^+$ standard solutions injected into it through the metal free injection valve (Rheodyne, model 7725) equipped with a 100 µL peek sample loop (Cheminert, VICI). Both solutions were delivered through a T-connector and peek tubing to the mixing coil made of 75 cm long teflon tubing of 1 mm internal diameter where the silver particles were synthesized and carried further into the flow through detection cell. The carrier and the reductant were pumped through the system at 0.3 mL/min flow rate each, what gave the total flow of the liquids equal to 0.6 mL/min and concentration of the reductant 0.3 mM after mixing with the sample. The silver colloids were produced in flowing solution of 0.6 mM $NaBH_4$ at pH 12.5 by injecting the silver ions of required concentrations (standards) into the carrier (double-deionized $H_2O$).

The TLS detection was performed by using a Krypton (Kr) laser (Innova 300C, Coherent) as the excitation beam (EB) source operating at 407 nm emission line and providing 115 – 195 mW output power at the location of the flow-through detection cell with a 1 cm optical pathway and 8 µL volume (Starna scientific, type 583.3.3). The EB was modulated with a frequency of 40 Hz by a mechanical chopper (Scientic instruments, Control unit model 300C, chopping head model 300CD, chopping disks model 300H). Temperature gradients, induced in the sample by the modulated EB, were probed by a He-Ne laser (Uniphase, Model 1103P) of 2 mW output power at output wavelength 632.8 nm (probe beam PB) and monitored by a pin photodiode (RBM - R. Braumann GmbH) placed behind a pinhole and an interference filter (Melles Griot, laser line filter, CWL 632,8 nm) combined with colored filter (Thorlabs 630nm Longpass), and connected to a lock-in amplifier (Stanford research instruments, Model SR830 DSP) and a PC. Both EB and PB beams were focused by a lens (25 mm diameter) of a focal distance 100 mm (EDMUND OPTICS, Bi-Convex, AR Coated: 350-700 nm) and directed onto the dichroic mirror (EDMUND OPTICS, 45° Reflective Dichroic Color Filters, reflection wavelengths 550-725 nm, transmission wavelengths 400-995 nm) by a set of mirrors (Edmund Optics, VIS, Precision Broadband Laser Mirror). The dichroic mirror cross overlapped the beams and directed them onto the flow through detection cell before reaching the detector.

The measurements were performed for the optimized experimental conditions. The optimal chemical condition were determined by the use of Yates' algorithm and the normal probability plot [21,32]. The maximum signal response (peak height), with the lowest standard deviation was achieved for 0.6 mM concentration of $NaBH_4$ and pH 12.5, whereas the optimal total flow rate resulting from carrier and the reductant was determined as 0.6 mL/min. Further optimization of the experimental setup was performed for the range of sample loops 10-500 µL and for the range of mixing coils 50-75 cm. It was found that a sample loop of 100 µL and a mixing coil of the length 75 cm provide the maximum peak height while maintaining the peak dispersion low. So, the optimized experimental conditions were: the flow rate = 0.3 mL/min for both the carrier and the reductant, injection loop = 100 µL, 0.6 mM concentration of $NaBH_4$ at pH 12.5, 75 cm long mixing coil of internal diameter 1 mm. This is also in agreement with the kinetics of nanosilver formation in a given system (Fig. 3), where maximum signal is approached at 225 s, corresponding to the residence time in a mixing coil at 0.6 mL/min flow rate. The $Ag^0$ was formed by $NaBH_4$ reduction in its flowing solution what eliminates the need of stabilizers or polymers which can interfere with the metal surface and influence the results of analysis [17].

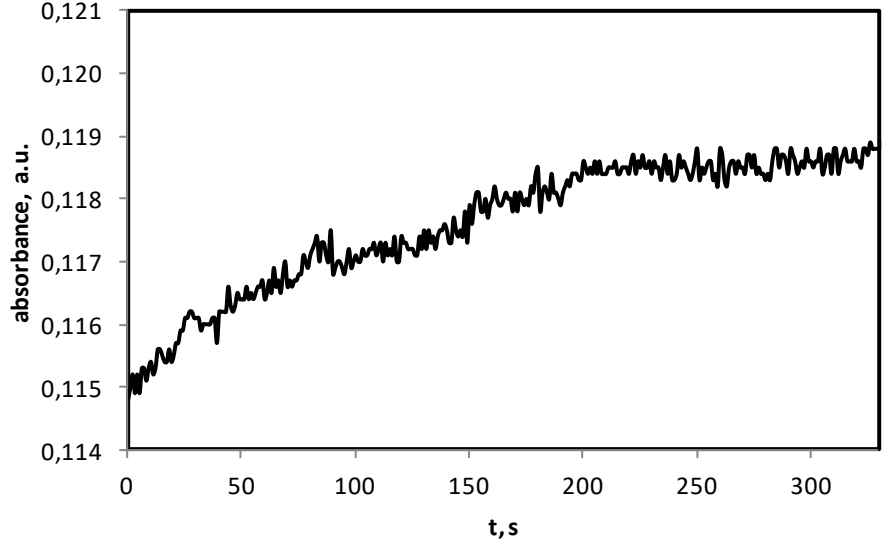

Fig. 3. Kinetics of the Ag$^+$ (1000 µg/L) reduction reaction by BH$_4^-$ as observed spectrophotometrically in a batch mode.

*3.3 Time resolved TLS*

The change in PB intensity at its centre introduced by TL can be written as [33,34]:

$$I(t) = I(0)\left[1 - \frac{\theta}{2}\tan^{-1}\left(\frac{2mV}{[(1+2m)^2+V^2](t_c/2t)+1+2m+V^2}\right)\right]^2 \quad (4)$$

where

$$\theta = -\frac{P_{th}Al(dn/dT)}{k\lambda_p} \quad (5)$$

and

$P_{th}$ - the EB power, $I$ - the sample thickness (cm), $l_p$ - the PB wavelength, $A$ - absorption coefficient of the sample, $dn/dT$ - refractive index gradient with temperature, t – time, $I(0)$ - the intensity of PB at $t = 0$, $k$ - thermal conductivity of the sample, $t_c$ - the characteristic time constant expressed as:

$$t_c = \frac{\omega_{e0}^2}{4D} \quad (6)$$

$\omega_{e0}$ - the radius of EB in the waist, $D$ - the thermal diffusivity of the Ag$^0$ NF.

$V$ is the ratio of the separation between the sample cell and the probe beam waist to the confocal distance of the probe beam described by the equation:

$$V = \frac{z_1}{z_c} + \frac{z_c}{z_2}\left[1 + \left(\frac{z_1}{z_c}\right)^2\right] \quad (7)$$

where $z_1$ - the distance of PB waist to the sample, $z_2$ – the distance from the sample to the detector, $z_c$ - the PB confocal (Rayleigh) distance, $m$ is the mode mismatch factor, which is described by the equation:

$$m = \left(\frac{\omega_{p1}}{\omega_{e0}}\right)^2 \quad (8)$$

$\omega_{p1}$ – the radius of PB, $\omega_{e0}$ – the radius of EB at the center of the sample.

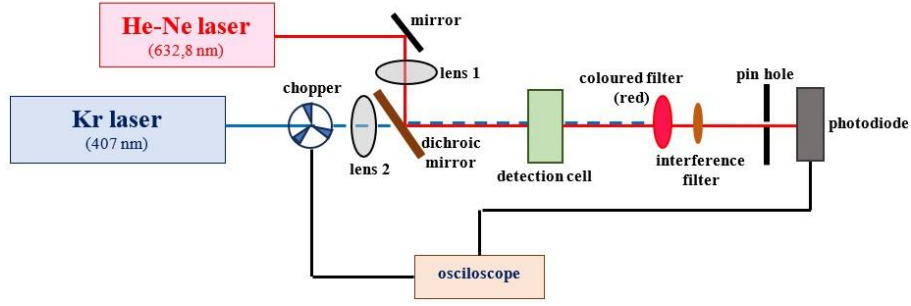

Fig. 4. Time resolved TLS experimental setup for Ag$^0$ NFs thermal diffusivity determination.

To determine the thermal diffusivity of Ag$^0$ NFs the TLS experimental setup was modified by removing FIA system and substituting with digital osciloscope (RIGOL, DHO900, 250 MHz bandwidth) as shown in Fig. 4.

### 3.4 Beam deflection spectrometry

The beam deflection spectrometry is based on sample illumination by a modulated excitation radiation. The absorbed EB energy is converted into heat, leading to temperature oscillations (TOs) in the examined sample and the fluid layer over its surface. These oscillations cause changes in the refractive index and its gradient, which are subsequently probed by another laser beam (PB). The interaction of PB with TOs causes its deflection of the PB trajectory from its initial direction of propagation according to the equation [35]:

$$\Delta z = n_0^2 s_T \int_0^\tau (\tau - \tau') \frac{\partial \vartheta_f}{\partial z} d\tau' \qquad (9)$$

where $n_0$ is the refracting index of fluid without induced TOs, $s_T = (1/n_0)(dn/dT)$ is the temperature coefficient of refractive index (thermal sensitivity), $\tau$ is the running complex coordinate along the PB trajectory. The PB is assumed to enter the experimental setup in the plane $z = 0$ and propagates in the positive direction of the OZ axis. The TOs $\vartheta_i$ in the $i$ layer of the analysed sample (see Fig. A1 and Supplement 1 ) fulfills the Fourier-Kirchhoff equation [36,37]:

$$D_{Ti}^{-1} \frac{\partial \vartheta_i}{\partial t} = \nabla^2 \vartheta_i + q_i k_{Ti}^{-1} \qquad (10)$$

where $q_i$ is internal heat sources power density, $k_{Ti}$ and $D_{Ti}$ are thermal conductivity and diffusivity, respectively. The EB incident light is modulated with frequency $f$ causing its intensity variation in the range of 0 to $I_0$. It is then absorbed by sample inducing internal heat sources in each layer of sample (see Fig. A1 and Supplement 1).

The deflection of PB trajectory leads to its amplitude change expressed as:
$$A(z) \cong A_0(1 + \Delta a) \qquad (11)$$
where $A_0$ is the amplitude of PB without the presence of TOs, $\Delta a$ is the correction introduced by PB deflection on refractive index gradients.

Another effect of PB interaction with TOs is its phase change expressed as [35]:

$$\Delta \Phi = k n_0^2 s_T \int_0^\tau \vartheta_f[z(\tau')] d\tau' \qquad (12)$$

Both the deflection and change in PB phase cause its intensity change that produces the photodeflection signal. In case of measurements by quadrant photodiode (QP), the BDS signal can be calculated as:

$$S_{BDS} = 2K_d \left\{ \int_0^{+\infty} - \int_{-z_0}^{0} dx \int_{-\infty}^{+\infty} dy [\text{Re}(\Delta a) - k\text{Im}(\Delta \Phi)] I_0 \right\} = A_{BDS}\cos(2\pi f t + \text{atan}(\Theta_{mI}/\Theta_{mR}) + \varphi_{BDS}) \quad (13)$$

$$\Delta a = \Delta a_1(z_{D1}) + \Delta a_2(z_{D2}), \qquad \Delta \Phi = \Delta \Phi_1(z_{D1}) + \Delta \Phi_2(z_{D2}) \quad (14)$$

where $K_d$ – the detector constant, $k$ – the PB wave number, $z_0$ – the height of PB over the sample surface, $A_{BDS}$ and $\varphi_{BDS}$ – the amplitude and phase of BDS signal, respectively.

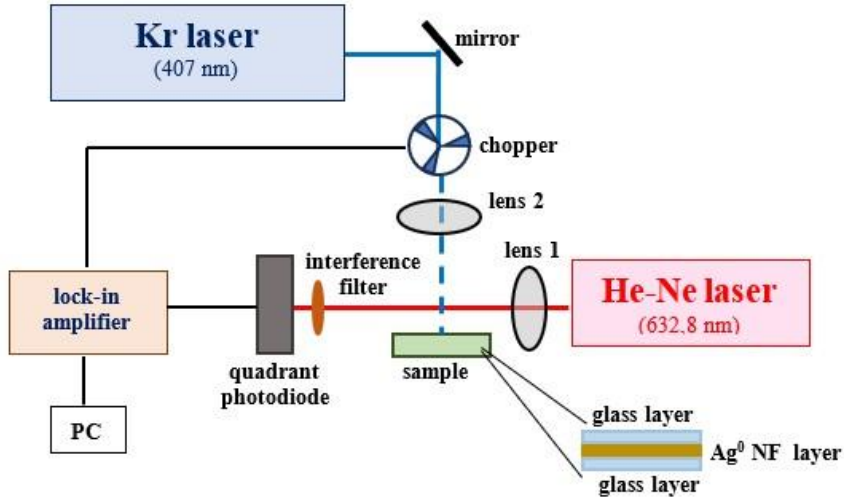

Fig. 5. BDS experimental setup for $Ag^0$ NFs thermal conductivity determination.

The source of EB was a Kr laser at 407 nm output wavelengh and 100 mW output power (Innova 300C, Coherent). Its intensity was modulated in the frequency range from 0.1 up to 1 Hz by variable-speed mechanical chopper (SCIENTIC INSTRUMENTS, Control unit model 300C, chopping head model 300CD, chopping disks model 300H). The obtained thermal diffusion length $\mu_{th} = \sqrt{D_s/\pi f}$ ($D_s$ thermal diffusivity of the sample) was of about 1.5 mm what enabled the information collection from all the sample volume that was placed on 3D translation stage to move it in x-y-z direction and optimize the experimental condition. EB was further directed on the sample prependicular to its surface by a set of mirrors (VIS, Precision Broadband Laser Mirror, EDMUND OPTICS) and formed by a lens of 100 mm focal length (Bi-Convex, AR Coated: 350-700 nm, EDMUND OPTICS) to the spot of about 2 mm². PB was a beam from a a He-Ne laser (Uniphase, Model 1103P) emitting 633 nm output wavelengh of 3 mW output power. It propagates parallel to the sample surface just skimming it. PB is focused by a lens of 40 mm (Bi-Convex, AR Coated: 350-700 nm, EDMUND OPTICS) to focus PB to a spot of around 50 μm. In such a way 1D configuration of the experimental setup is provided (PB size << EB size in the area of TOs) [35]. The BDS signal is collected by a position sensing detector (Quadrant Detector Sensor Head, 400 to 1050 nm, Thorlabs) equipped with an interference filter (laser line filter, CWL = 633 nm, Thorlabs) and connected to a lock-in amplifier (SRS830 DSP, Stanford Research Systems) and PC where the amplitude of BDS signal is recorded and processed by the use of Matlab software. The experimental setup is shown in Fig. 5.

*3.5 Spectrophotometry*

The plasmon absorption spectra of formed silver nanoparticles were recorded on the Lambda 650 PerkinElmer dual beam spectrophotometer in a 10 mm quartz cuvette (Hellma, spectral range 200-2500 nm, pathlength 10 mm, chamber volume 700 µL). All the spectra were recorded with respect to the double deionized water as a blank. UV-Vis spectrum of a silver free aqueous solution containing only $NaBH_4$ and the spectrum of borohydride free solution containing only $AgNO_3$ were also recorded to be subtracted from the spectral curves of silver samples to obtain UV-Vis spectra corresponding only to silver NPs.

*3.6 IonPac Cryptand G1 column*

The presence of interfering ions in NF solution may influence the value of determined NPs concentration, thus, properties of NF as advanced class of heat transfer fluid and/or disinfection agent. Because of that it is important to develop a procedure for selective detection of silver compounds in the presence of interfering ions. The solution can be found by implementation of IonPac Cryptand G1 column (30mm×3mm i.d., Dionex) in FIA-TLS detection system.

Being a bi-cyclic ligands for a variety of cations, 2.2.2 cryptand is capable of binding cations using both <u>nitrogen</u> and <u>oxygen</u> donor groups. The complexing of ions confers size-selectivity and thus enables discrimination among cations. The binding constants $K$ of 2.2.2 cryptand with different ions are listed in Table 1. Since the value of $K$ for $Ag^+$ is of the highest ($10^{9.6}$) of all ions expected in water supplies, which were demonstrated to interfere in Ag detection, the formation of a complex with $Ag^+$ is highly favored and hence by the selective retention of the $Ag^+$ on the IonPac Cryptand G1 column, over other interfering ions, selective determination of silver compound is expected.

The procedure consists of three steps that are as follows:
1. the column was loaded with 10 mL of 1 $mgL^{-1}$ $Ag^+$ ($AgNO_3$) at flow rate of 0.5 $mLmin^{-1}$. Eluent received from the column was collected and injected into the FIA system with TLS detection to evaluate the $Ag^+$ amount unretained by reduction $Ag^+$ to $Ag^0$ with $BH_4^-$,
2. the column was washed with 1 mL of $H_2O$ with 1 $mLmin^{-1}$ flow rate to remove the solution left inside after loading it with 1 $mgL^{-1}$ of $Ag^+$. Finally, to recover the retained $Ag^+$ by forming the $Ag(NH_3)_2^+$ complexes with ammonia, the 10 mL of $NH_4^+$ ($NH_4OH$) at pH 11 as eluent was passed at a flow rate of 0.5 $mLmin^{-1}$ through the column without its connection to the system,
3. To dissolve the complexes and bring silver back to its ionic form, a 20 µL of concentrated $HNO_3$ was added to the collected eluate which was further analyzed with FIA coupled to TLS unit,
4. the column was washed with 0.1 M of $KNO_3$ at 0.5 $mLmin^{-1}$ flow rate for 2 h and loaded with 10 mL of $H_2O$ with 1 $mLmin^{-1}$ for storage.

**Table 1. Binding constants $K$ of 2,2,2 cryptand with different ions [38].**

| Ions | log$K$ |
|---|---|
| $Li^+$ | ~1 |
| $Cs^+$ | <2 |
| $Na^+$ | 3.9 |
| $Ca^{2+}$ | 4.4 |
| $NH_4^+$ | 4.5 |
| $K^+$ | 5.4 |
| $Cu^{2+}$ | 6.8 |
| $Ba^{2+}$ | 9.5 |
| $Ag^+$ | 9.6 |

*3.7 Scanning electron microscopy*

A scanning electron microscope (SEM, JSM-7100F, JEOL) was used to collect the micrographs of synthesized $Ag^0$ NPs. It was coupled to energy dispersive X-ray (EDX) detector (XMax-80, Oxford) working at primary beam voltage 20.0 kV. The SEM images were recorded in secondary electron and back-scattered electron detection working modes.

## 4. Results and discussion

*4.1 Morphology*

The morphology of the synthesized $Ag^0$ NPs at different silver concnetrations was investigated by SEM. The recorded SEM micrographs are presented in Fig. 6. It is observed that nanoparticles (NPs) with sizes up to 40 nm are formed, while larger $Ag^0$ nanoparticles tend to aggregate, resulting in clusters with diameters of up to 700 nm (Fig. 7).

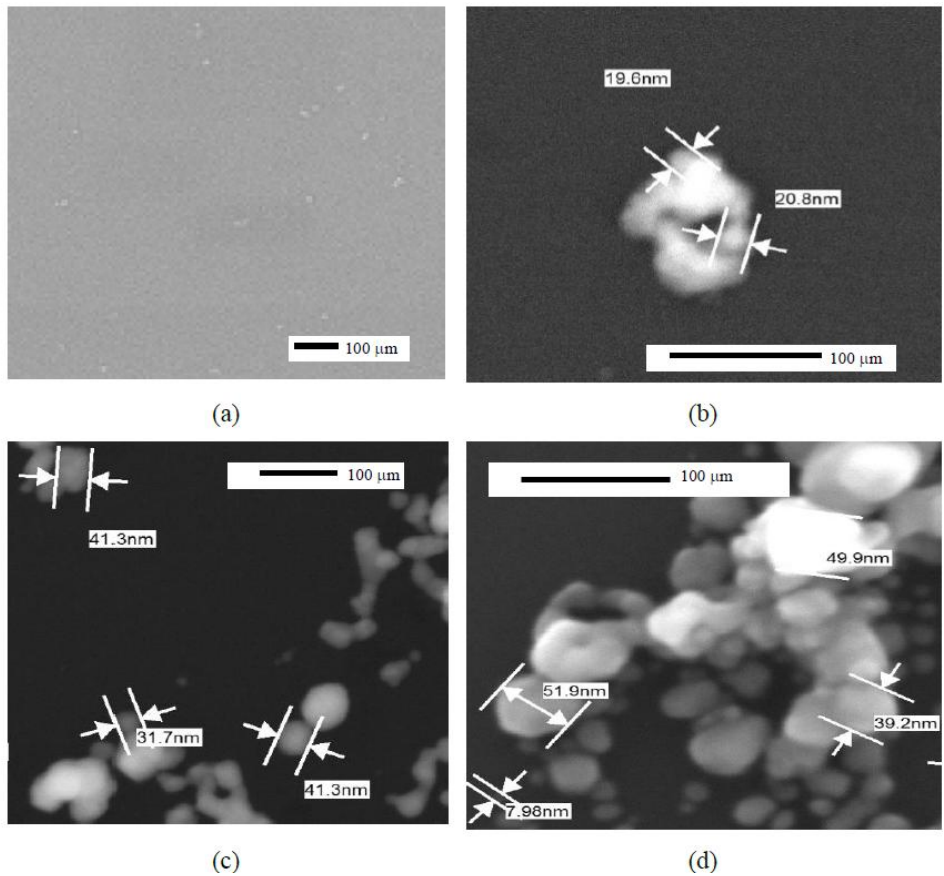

Fig. 6. SEM images of Ag$^0$ NPs at different concentrations (a) 50 mg/L; (b) 1 mg/L; (c) 5 mg/L; (d) 15 mg/L.

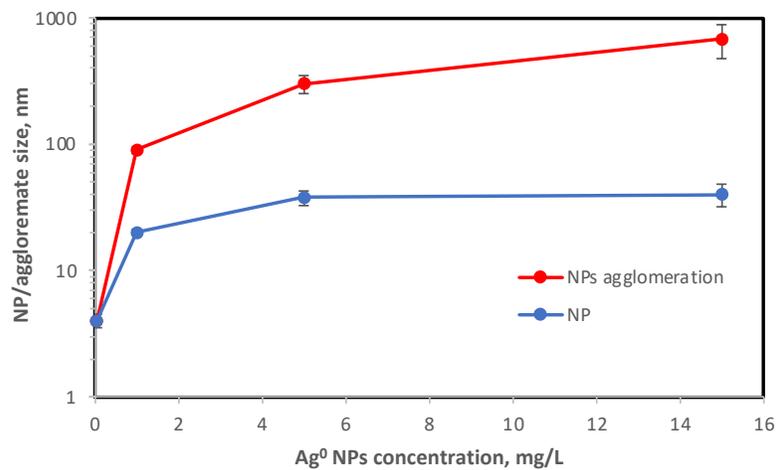

Fig. 7. Ag$^0$ NPs and their agglomeration size as a function of Ag$^0$ NPs concentration.

*4.2 Absorption spectra*

The absorption spectra of silver nanoparticles in aqueous solution exhibit a plasmon resonance peak located in the visible spectral range at about 400 nm (Fig. 8), which results from the interaction of an optical field with silver nanoparticles and production of the electron gas oscillation (Fig. 9). Optical field (OF) is a vectorial field that is combined of coupled electric and magnetic fields described by the Maxwell's equations [39]. OF determines an optical wave (OW) in time and space. As a result of interaction with metallic NPs the electric field of incident electromagnetic (E-M) radiation penetrates the metal polarizing the conduction electrons and inducing plasmons that are oscillations of free electrons in metal NPs that have a form of a negatively charged electron cloud (EC) coherently displaced from its equilibrium position around a positively charged lattice. In such a way EC harmonic oscillation are transverse magnetic (TM) in nature and are induced in the direction parallel to the electric field of incident E-M radiation [40,41]. It must be stated that only E-M radiation of frequency being in resonance with these EC oscillations are capable of exciting surface plasmon resonance (SPR). SPR occurs along interface of two media with opposite sign of dielectric permittivity. The EC movement creates an E-M field which is added to incident one what leads to the field enhancement within the NPs as well as in its vicinity at specific wavelength what is seen as resonances in NPs absorption spectra reflecting the enhancements in both NPs optical scattering and absorption cross-sections [42].

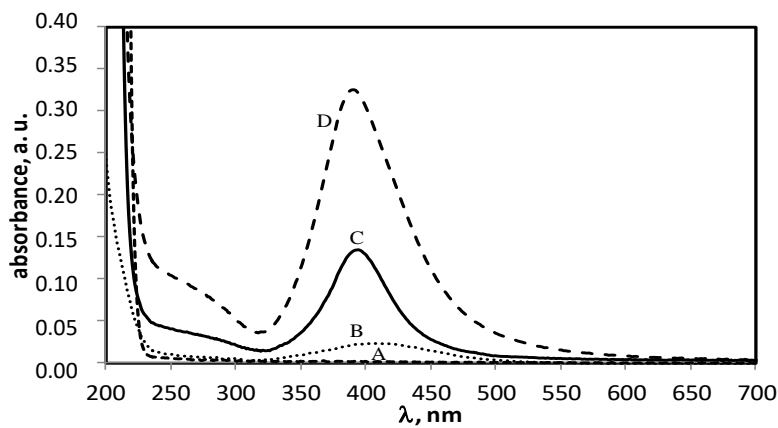

Fig. 8. UV-vis spectra of silver colloids corresponding to 1000 (C), 3000 µg/L (D) silver nitrate concentration used during the $Ag^0$ synthesis, as well as spectra of $NaBH_4$ solution (A) used as a reductant and $AgNO_3$ aqueous solution (B). All the spectra were recorded with water as the reference.

The silver nanoparticles' absorption peak increases in height with the increase in the silver ion concentration. The 400 nm peak for 1000 and 3000 µg/L appears without any shoulder indicating the absence of agglomeration. The formation of aggregates is a random process which is caused by the action of short distance van der Waals forces responsible for particle aggregation and conservation of the local anisotropy of the environment. The peak at 400 nm is the surface plasmon band which corresponds to the average spherical particle's with diameter between 10-30 nm.

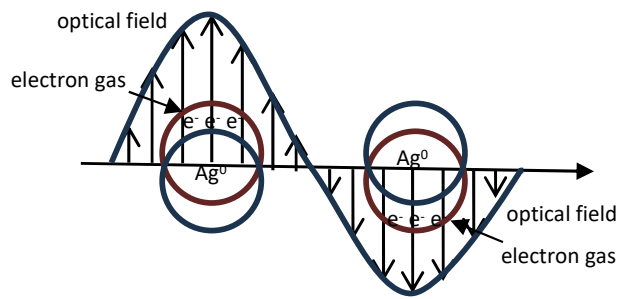

Fig. 9. Surface plasmon resonance in which the free electrons in $Ag^0$ NP are induced into oscillation as a result of NP interaction with a wavelength of excitation radiation.

The reactants, used to prepare silver colloids, e.g. silver nitrate and sodium borohydride, exhibit strong absorption in the UV region of the spectrum (Fig. 8) what is expressed on silver colloids absorption spectra as additional peak at 190 – 250 nm. Additionally, the absorption spectra of silver ions show a small peak at 400 nm which corresponds to silver hydroxide which are formed when water at pH 12.5 is added into the $Ag^+$ solution instead of the reductant.

### *4.3 LOD of the method*

For the FIA-TLS calibration curve, the value of the TLS signal is plotted as a function of the concentration of injected standard solutions (Fig. 10.). In general, this relationship is not linear (Eq.1), and depends on the absorbance of the sample. But for low absorbances (A < 0.1) linear approximation is satisfactory [38,43]. In our experiments the values of TLS signal were determined as peak heights average from triplicate injections of samples or standard silver solutions in water. For the investigated concentrations (< 600 µg/L) a linear relation between $\Delta I/I$ and silver ion concentration was observed (Fig. 10) and approximated by a linear equation following least-squares regression procedures. This enabled determination of the intercept and slope of the regression line and calculation of the LOD of the method for different excitation powers.

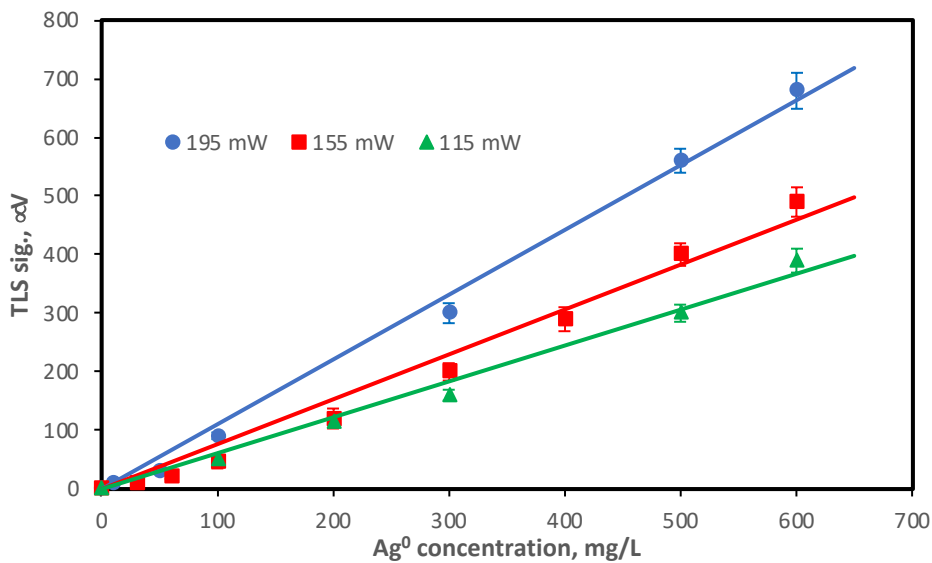

Fig. 10. The calibration curves for three values of output power of the Krypton laser (407 nm) used for the excitation.

It is seen from Fig. 10 that the linearity range of the method is between 20 and 600 µg/L. Good reproducibility of the method is demonstrated by the standard deviation of less than 2.5 % achieved for four replicate measurements for each sample. The LODs were calculated for the signal to noise ratio of 3 and for three values of excitation beam power of the 407 nm emission line, which is close to the resonance band of produced silver colloids. Slopes are in agreement with the linear dependence of TLS signal on the excitation power (less than 6% lower than predicted). The calculated LOD values were: 1.2 µg/L for the laser power of 115 mW ($y = 644.78x – 14.99$; $r^2 = 0.992$), 0.8 µg/L for the laser power of 155 mW ($y = 844.66x – 20.75$; $r^2 = 0.991$) and 0.7 µg/L for the laser power of 195 mW ($y = 1156.41x – 13.61$; $r^2 = 0.998$). All parameters used for calculation of LODs are presented in Table 2.

**Table 2. Results of the LOD determination**

| P, mW | slope of the calibration curve, µV/ µgL$^{-1}$ | standard deviation of the blank, µV | LOD, µg/L |
|---|---|---|---|
| 115 | 645 ± 32 | 0.246 | 1.17 ± 0.06 |
| 155 | 845 ± 74 | 0.216 | 0.85 ± 0.06 |
| 195 | 1056 ± 40 | 0.241 | 0.68 ± 0.03 |

It is seen that by increasing EB power, if the standard deviation of the blank does not increase proportionally with its increase, the sensitivity of the method can be also increased. The achieved LODs of the method are sufficiently lower even than the maximum contaminant level (MCL) for silver in water, which varies from 25 µg/L to 75 µg/L in different countries and 200-500 µg/L in case of space missions and also much lower than LOD achieved by spectrophotometry which was 50 µg/L [21].

*4.4 Detection of silver colloids in the presence of interfering ions*

As shown previously [21], the method of Ag colloids detection in the presence of contaminants was already tested. The analysis was performed for the level of foreign ions not interfering $Ag^0$ determination. In this paper, the analysis was performed for such a range of contaminants for which their effect on Ag detection is observed. To determine the amount of $Ag^+$ absorbed on the column, loaded previously with 10 mL aliquots of 100, 250, 1000 µg/L $Ag^+$ working solutions, the recovery procedure was performed using $NH_4^+$ solution at pH 11 as an eluent (see section 3.6). The obtained recovery, expressed as a percentage of $Ag^+$ amount present in initial working solutions, was 80±3% for $Ag^+$ concentration range from 100 to 1000 µg/L of $Ag^+$ (Fig. 11).

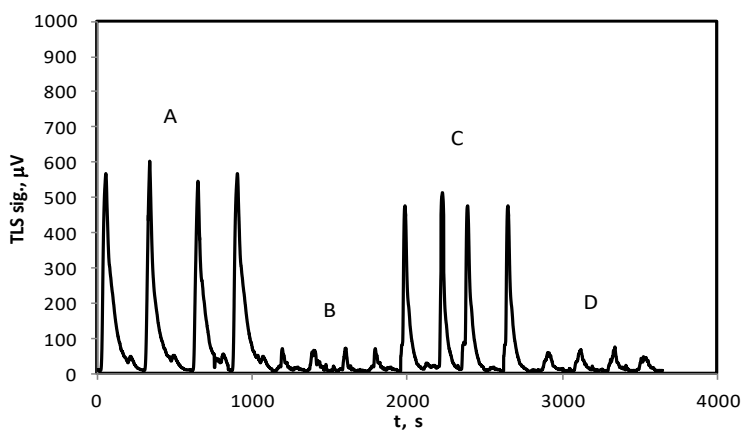

Fig. 11. TLS signal for on-line production of colloid silver in: A – 1000 µg/L of $Ag^+$ standard solution; B – eluent collected from the IonPac Cryptand G1 column after loading it with 1000 µg/L of $Ag^+$; C – eluent collected by washing the IonPac Cryptand G1 column (loaded with 1000 µg/L of $Ag^+$) with $NH_4^+$ solution at pH 11 (20 µL of $HNO_3$ added to eluent); D – blank: $NH_4^+$ solution at pH 11 with 20 µL of $HNO_3$ added.

In the next step the recovering procedure was repeated for solutions of interfering metal ions that were found to contribute to the detection of $Ag^+$ in water supplies (e.g. $Fe^{3+}$, $Mn^{2+}$, $Cu^{2+}$) if present in the solution in the amount of their MCL. In the procedure the concentrations of interfering ions, loaded first in the column, were taken to be 0.3, 3, 30 ($Fe^{3+}$); 2, 20, 200 ($Mn^{2+}$) and 0.8, 8, 80 ($Cu^{2+}$) times higher than their MCL in water. It was found that for some ions the retention on the column was as high as in case of $Ag^+$ (e.g. $Fe^{3+}$; $Cu^{2+}$). However, in this case the elution performed using $NH_4^+$ solution (pH 11) reached the level of less than 30% during the first column washing (Fig. 12, Table 3) what presents relatively low value when compared to the case of $Ag^+$ for which the recovery was found to be 80%.

Another group of investigated ions, were ions weakly trapped by the column (e.g. $Mn^{2+}$) (Fig. 13). Thus, their amount in the eluent obtained by washing the column was only about 7% of the initially present concentration.

The results of the analysis are presented in Table 3. It can be concluded that the influence of all tested interfering ions on the Ag detection by the use of IonPac Cryptand G1 column is not expected to be significant, especially because of the fact that their MCLs are much lower than the levels of tested concentrations.

**Table 3. Results of the analysis of washing out the trapped ions from the IonPac Cryptand G1 column by $NH_4^+$ at pH 11.**

| Ions | conc., µgL$^{-1}$ | TLS sig., µV | fraction of the initial TLS value received after washing it with $NH_4^+$ at pH 11, % |
|------|-------------------|--------------|---------------------------------------------------------------------------------------|

| | | | 1st | 2nd | 3rd |
|---|---|---|---|---|---|
| $Ag^+$ | 1000 | 531 ± 17 | 82.0 ± 3.2 | 12.0 ± 0.3 | 8.0 ± 0.3 |
| $Fe^{3+}$ | 4000 | 579 ± 18 | 27.0 ± 0.8 | 15.0 ± 0.4 | 13.0 ± 0.4 |
| $Cu^{2+}$ | 5000 | 308 ± 12 | 28.0 ± 0.9 | 17.0 ± 0.6 | 11.0 ± 0.3 |
| $Mn^{2+}$ | 3000 | 414 ± 14 | 7.0 ± 0.2 | 3.00 ± 0.01 | 3.00 ± 0.01 |

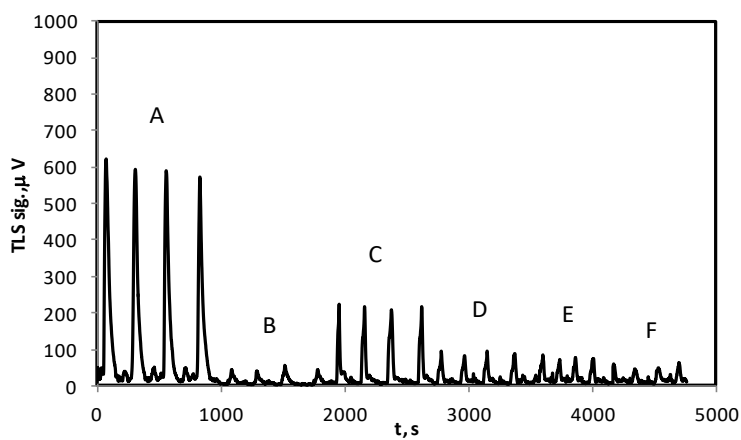

Fig. 12. TLS for: A – 4000 µg/L of $Fe^{3+}$; B – eluent collected from the IonPac Cryptand G1 column after loading it with 4000 µg/L of $Fe^{3+}$; C – eluent collected after 1st washing the IonPac Cryptand G1 column with $NH_4^+$ at pH 11 and adding to it 20 µL of $HNO_3$; D – eluent collected after 2nd washing the IonPac Cryptand G1 column with $NH_4^+$ at pH 11 and adding to it 20 µL of $HNO_3$; E – eluent collected after 3rd washing the IonPac Cryptand G1 column with $NH_4^+$ at pH 11 and adding to it 20 µL of $HNO_3$; F – $NH_4^+$ at pH 11 with 20 µL of $HNO_3$ added.

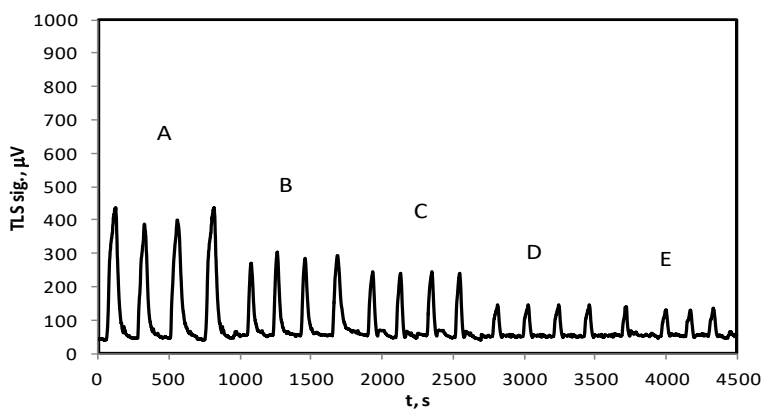

Fig. 13. TLS signal for: A – 3000 µg/L of $Mn^{2+}$; B – eluent collected from the IonPac Cryptand G1 column after loading it with 3000 µg/L of $Mn^{2+}$; C – eluent collected after 1st washing the IonPac Cryptand G1 column with $NH_4^+$ at pH 11 and adding to it 20 µL of $HNO_3$; D – eluent collected after 2nd washing the IonPac Cryptand G1 column with $NH_4^+$ at pH 11 and adding to it 20 µL of $HNO_3$; E – $NH_4^+$ at pH 11 with 20 µL of $HNO_3$ added.

To confirm these expectations, the retention procedure on the Cryptand column was repeated for the solution of 100 µg/L $Ag^+$ containing the interfering ions $Fe^{3+}$, $Mn^{2+}$ and $Cu^{2+}$ at the level of their MCL that are 300 µg/L, 50 µg/L and 1300 µg/L, respectively. The analysis was performed three times for every single interferent separately. On the basis of the value of the recovery for $Ag^+$ found previously (i.e. 80±3%), the value of determined $Ag^+$ concentration was found for all tested foreign ions. The value of $Ag^+$ wash out from the column was found by the use of calibration curve. Taking into account the value of the recovery (80±3%) the determined values of $Ag^+$ were: 107, 104 and 120 µg/L in case of added $Fe^{3+}$, $Mn^{2+}$ and $Cu^{2+}$ respectively what differs from the real value (100 µg/L) by not more than 20%. This indicates the contribution of the interferents to the determined value of silver ions to be at the level of 7%, 4% and 20% for $Fe^{3+}$, $Mn^{2+}$ and $Cu^{2+}$ as interfering ions, respectively. Because of that, the presented method seems to be suitable for determining the concentration of $Ag^+$ in the presence of other ions in water environments as rivers, lakes or tap water.

To verify this statement, the analysis was performed again for the case of 100 µg/L $Ag^+$ solution containing all the listed above interferents together in the amounts as mentioned before (107 µg/L of $Fe^{3+}$, 104 µg/L of $Mn^{2+}$, 120 µg/L of $Cu^{2+}$) (Fig. 14). The determined values of $Ag^+$ in this case were: 122 µg/L. It means that contribution of all interfering ions added together to $Ag^+$ solution, to the determined value of silver $Ag^+$, was found to be 22%.

It needs to be noted that the analysis was performed for the concentration of interferents at their MCL. Their average amounts in water supplies are: 46 µg/L [44], 13 µg/L [45] and 10 µg/L [46] in case of $Fe^{3+}$, $Mn^{2+}$ and $Cu^{2+}$, respectively. These values are 5 – 13 times lower that those tested in the experiment, so their influence on the $Ag^+$ determination in real samples would be even less than achieved in our experiment. Taking that into account it can be said that the IonPac Cryptand G1 column is a useful tool for determining the concentration of $Ag^+$ ions in the presence of contaminants in water supplies because by the use of it the influence of interfering ions on the determined value of $Ag^+$ can be successfully eliminated.

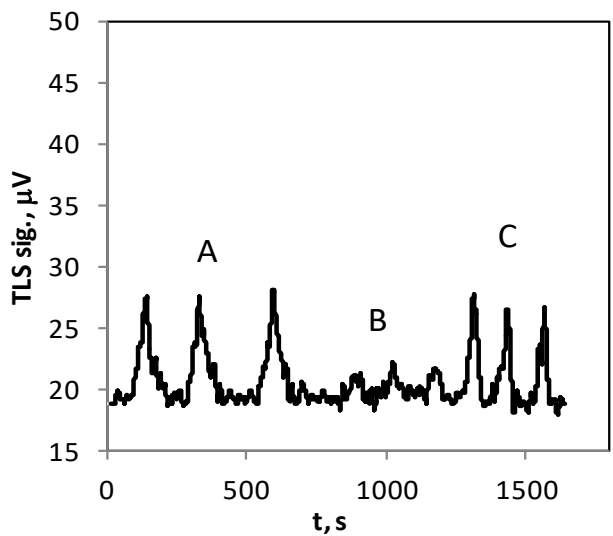

Fig. 14. TLS signal for: A – 100 µg/L of $Ag^+$ containing $Fe^{3+}$, $Mn^{2+}$ and $Cu^{2+}$ at the level of their MCL; B – eluent collected from the IonPac Cryptand G1 column after loading it with 100 µg/L of $Ag^+$ containing $Fe^{3+}$, $Mn^{2+}$ and $Cu^{2+}$ at the level of their MCL; C – eluent collected after 1st washing the IonPac Cryptand G1 column with $NH_4^+$ at pH 11 and adding to it 20 µL of $HNO_3$.

An option for the future is to use the ethylenediaminetetraacetic acid (EDTA) as an eluent which shows strong ability to complex metal ions such as $Fe^{3+}$, $Mn^{2+}$ or $Cu^{2+}$ and weak for $Ag^+$. Its binding constants (log$K$) [21,47,48] for mentioned ions are presented in Table 4. It is expected that using EDTA as an eluent we are able to significantly improve the procedure of $Ag^+$ determination by the use of IonPac Cryptand G1 column.

Table 4. Binding constants *K* of EDTA with different ions

| Ions | log*K* |
|---|---|
| $Ag^+$ | 7.32 |
| $Mn^{2+}$ | 14.0 |
| $Cu^{2+}$ | 18.9 |
| $Fe^{3+}$ | 25.2 |

*4.5 Thermal diffusivity determination*

The EB is absorbed by the $Ag^0$ NPs conductive electrons inducing plasmonic resonance oscillations what in turn generates the hot electrons that further relax non-radiatively by emission of phonons causing the lattice rapid heating. Next, the heat is exchanged between the phonons and the base fluid what leads to creation of thermal gradient and further variation in refractive index of the whole medium what induced TL of concave shape ($dn/dT < 0$). The time dependent thermal lens signal for different concentration of silver NFs is shown in Fig. 15. The thermal diffusivities of $Ag^0$ NFs with different silver concentrations were found performing fitting of theoretical TLS signal (Eq. 4) to the measured data (Fig. 15). The obtained values of thermal diffusivities are presented on Table 5.

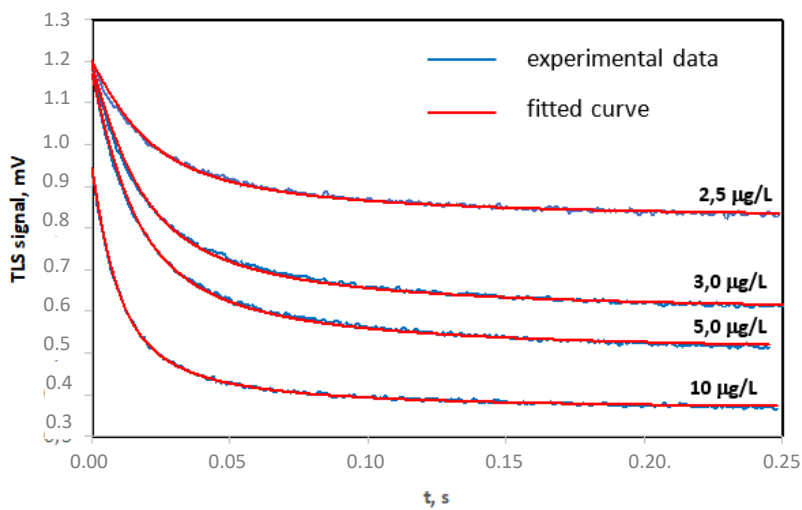

Fig. 15. Time dependence of TLS signal of silver NFs containing different $Ag^0$ concentration (experimental data together with fitted theoretical curves).

Table 5. Values of thermal diffusivities of silver NFs with different $Ag^0$ concentrations obtained by TLS.

| $Ag^0$ concentration, mg/L | Water | 2.5 | 3.0 | 4.0 | 5.0 | 6.0 | 8.0 | 10 | 50 | 100 |
|---|---|---|---|---|---|---|---|---|---|---|
| Thermal diffusivity, cm$^2$/s | 0.145 ± 0.006 | 0.186 ± 0.008 | 0.202 ± 0.009 | 0.208 ± 0.009 | 0.222 ± 0.010 | 0.238 ± 0.010 | 0.288 ± 0.012 | 0.362 ± 0.016 | 0.149 ± 0.006 | 0.151 ± 0.006 |
| Ratio of $Ag^0$ NF/water thermal diffusivity | - | 1.283 ± 0.044 | 1.393 ± 0.045 | 1.435 ± 0.044 | 1.531 ± 0.046 | 1.638 ± 0.043 | 1.986 ± 0.047 | 2.497 ± 0.047 | 1.028 ± 0.041 | 1.041 ± 0.040 |

It is seen that the addition of $Ag^0$ NPs influences the thermal diffusivities of the base fluid by improving them for $Ag^0$ NPs concentration up to 10 mg/L. It means that with the increse of NPs concentration the thermal diffusivity ration of NF/base fluid also increases. The increase in the NPs volume fraction effectively decreases the NF's volumetric specific heat [49–52], thus, the thermal diffusivity of the whole solution. This implies that for inducing the same temperature change, NF requires more amount of energy than the base fluid. This is because of the enhanced thermal conduction in the NF. When the nanoparticle concentration increases in a NF, the specific heat capacity of the fluid decreases. This is because of - (i) the lower specific heat capacity of the nanoparticle with respect to the base fluid. The increase of the nanoparticle hence lowers the heat capacity of the mixture. (ii) the interaction of nanoparticle-base fluid also can affect the heat capacity due to enhanced phonon transport or interfacial energy effects.

With further increase in $Ag^0$ concentration (over 10 mg/L) the NF thermal diffusivity decreases and stabilizes abround 0.15 cm$^2$/s. The reason for that is the effect of NPs aggregation what decreases the effective particle surface area-to-volume ratio increasing the NF's specific heat what consequently leads to decrease in its thermal diffusivity.

The NF's thermal properties are influenced by the chemical environment in which the NPs are suspended. In our case the $Ag^0$ are surrounded by $BH_4^-$ ions, which prevents NPs from aggregation. It means that around the NP a layer is formed that modifies the NP surface–fluid interface. In NFs the heat is conducted through the interface between the NP and base fluid [53]. Consequently, an increase in the interfacial area results in the increase in the rate of heat conduction what also enhances the NF thermal diffusivity [53,54].

*4.6 Thermal conductivity determination*

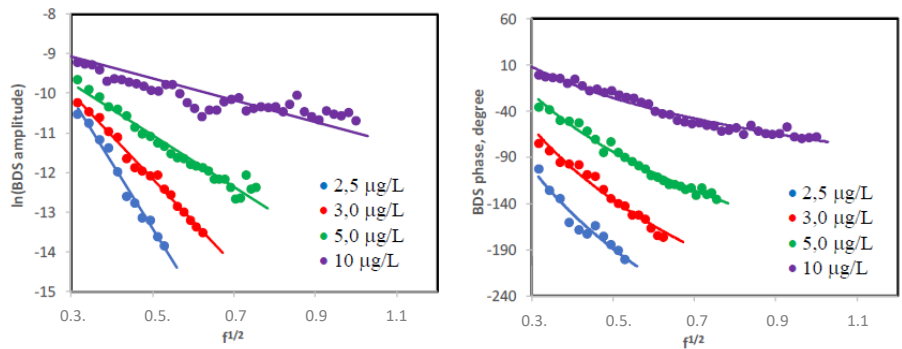

Fig. 16. BDS signal dependence on the square root of the EB modulation frequency.

The BDS signal collected from the Ag$^0$ NFs placed in cell closed with a cover of 1.25 mm thickness is relatively noisy (Fig. 16). Thus, the height of PB over the sample's surface is about 1.30 mm. For such a height the induced TOs, in the air over the sample surface, are weak what significantly decreased the value of BDS signal. Also, the BDS signal strongly decreases with the increase in modulation frequency of EB. All of these make the BDS measurements difficult to perform for a sample configuration introduced in the work (Supplement 1) and additional noise is introduced in the BDS measurements as seen in Fig. 16. Furthermore, she BDS signal depends on both the geometrical parameters of the experimental setup (e.g. height of the probe beam over the sample) and the properties of the examined samples (e.g. its thermal diffusivity and conductivity) (Supplement 1). To increase the sensitivity of the analysis, the geometrical properties of the experimental setup are determined for the sample with known properties and kept constant while performing analysis of Ag$^0$ NFs while the thermal diffusivities of Ag$^0$ NFs are found using TLS. In such a way the number of unknown properties in BDS measurement are limited only to one parameter that is the thermal conductivity of Ag$^0$ NF what increases the sensitivity of the thermal conductivity determination. The sensitivity of BDS technique for NF's thermal conductivity determination was defined as LOD of Ag$^0$ concentration in NF for which the thermal conductivity can be distinguished from the its value for blank signal (water). The calculated LOD was 1.25 mg/L.

It is seen from Table 6 that the thermal conductivity of Ag$^0$ NFs rises with the Ag$^0$ concentration to a maximum value of 1,45 W/mK at a concentration of about 10 mg/L. In this case the NF's thermal conductivity is over 140% higher than that one of pure water (Fig. 16). At higher Ag$^0$ concentration the NFs thermal conductivity decreases since the NPs tend to aggregate creating greater particles of mm size [55] as a result of what the concentration of small-scale Ag$^0$ NPs suspended in water decreases. Furthermore, larger particles have smaller area-to-volume ratio and thus, worse thermal properties comparing to smaller particles. Consequently, at high concentrations of Ag$^0$ NPs, the particles tend to stick together forming greater particles what decreases the thermal properties of NF with respect to the optimal concentration. The NPs deposition is another phenomena that is present in Ag$^0$ NF of higher concentrations that occurs due to the process of NPs aggregation. As a result of NPs deposition, the thermal resistance increased what deteriorates the heat transfer on the NF. NF is a two-phase fluid in which NPs undergo the Brownian motion. The random motion of NPs is directly temperature dependent and inversely proportional to the NP's size. Thus, Brownian motion introduces additional mechanism of convective heat transfer [56–58]. The force of molecular collision between molecules of water and suspended NPs is larger in case of smaller NPs than for larger ones and increases with the increase in the rate of temperature rise. In means that the heat tranfer rate can be increased by intensifying the NPs Brownian motion [59–61].

**Table 6. Values of thermal conductivities of silver NFs with different Ag$^0$ concentrations obtained by BDS.**

| Ag$^0$ concentration, mg/L | Water | 2.5 | 3.0 | 4.0 | 5.0 | 6.0 | 8.0 | 10 | 50 | 100 |
|---|---|---|---|---|---|---|---|---|---|---|
| Thermal conductivity, W/mK | 0.60 ± 0.03 | 0.76 ± 0.03 | 0.82 ± 0.03 | 0.84 ± 0.03 | 0.89 ± 0.04 | 0.95 ± 0.04 | 1.15 ± 0.05 | 1.45 ± 0.06 | 0.62 ± 0.03 | 0.63 ± 0.03 |
| Ratio of Ag$^0$ NF/water thermal conductivity | - | 1.267 ± 0.055 | 1.367 ± 0.055 | 1.394 ± 0.055 | 1.483 ± 0.063 | 1.589 ± 0.063 | 1.927 ± 0.072 | 2.417 ± 0.084 | 1.023 ± 0.058 | 1.034 ± 0.057 |

## 5. Discussion

The obtained results indicate that both the thermal diffusivity and conductivity of $Ag^0$ NFs increase with an increase in $Ag^0$ NPs for lower concentrations up to 10 mg/L of $Ag^0$ in water. Furthermore, the addition of $Ag^0$ NPs to water provides up to 150 % increase in the NF's thermal diffusivity and up to 140 % increase in its thermal conductivity (Fig. 17.). It can be stated that the presence of $Ag^0$ NPs leads to thermal properties enhancement, thus positively contributes to the rise in the heat transfer rate of the whole liquid for lower voulme fraction of $Ag^0$ NPs.

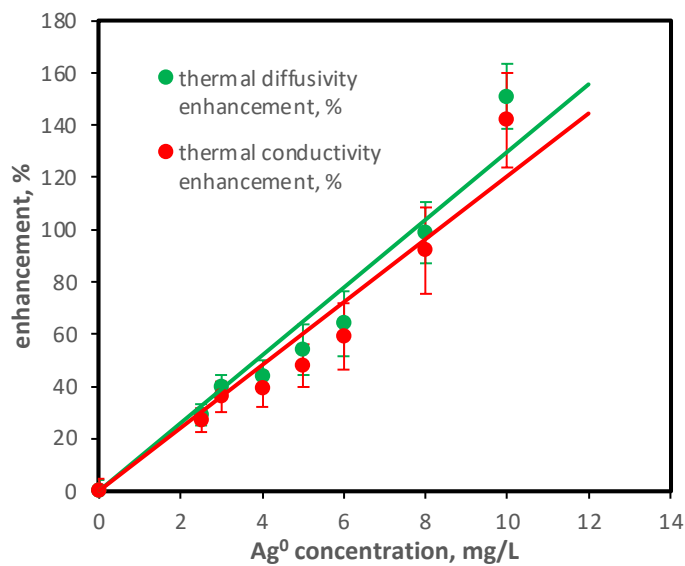

Fig. 17. The enhancement in thermal properties of $Ag^0$ NFs as a function of NPs concentration.

## 6. Conclusions

This study presents the possibility of highly sensitive detection of colloidal silver in water by in a FIA system and its detection by dual beam TLS. The achieved LOD is much lower than MCL for silver in water. It means that FIA with TLS detection is a sensitive and simple method for monitoring the quality of water with LODs about 30 times lower than in case of spectrophotometric measurements. The sensitivity of TLS technique can be increased by increasing the laser power of the pump beam and changing its wavelength to such a range for which the absorption of blank (0.3 mM $NaBH_4$) is lower but higher for the signal. In such a situation the increase of laser power does not increase the signal-to-noise ratio but improves the LOD but the increase in LODs is also limited by the nonlinearity effects of TLS signal behavior for high absorption of the species. The received LODs of $Ag^0$ detection were: 1.5 µg/L for 115 mW of the excitation radiation and 0.8 µg/L and 0.7 µg/L for 155 mW and 195 mW respectively. It was also demonstrated that several foreign ions, in amounts normally presented in water, interfere strongly in formation of $Ag^0$, thus its antibacterial efficiency, so that fact should be taken into account when using it as disinfection agent. It is possible to optimize this technique by using the IonPac Cryptand G1 column that ensures selective detection of silver in the presence of foreign ions. Due to that it is possible to implement this technique for determination of different compounds in environmental samples.

Furthermore, the colloid silver nanofluids containing different amounts of $Ag^0$ NPs were found to have increased thermal properties for lower concentrations up to 10 mg/L and provided the thermal diffusivity and conductivity enhancement at the level of 150 and 140 % for thermal diffusivity and conductivity, respectively, when compared with the properties of water as base fluid. Thus, the thermal enhancement of such NFs contributed to higher rise in its thermal performance over the water as base fluid.

## 8 Back matter
### 8.1 Funding
**Funding.** Slovene Research and Innovation Agency through Grant No. P2-0393
### 8.2 Acknowledgment
**Acknowledgment.** Financial supports were provided for D. Korte and M. S. Swapna from Slovene Research and Innovation Agency through Grant No. P2-0393; Advanced materials for low-carbon and sustainable society.
### 8.3 Disclosures
"The authors declare no conflicts of interest."
### 8.4 Data availability statement
**Data availability.** Data underlying the results presented in this paper are not publicly available at this time but may be obtained from the authors upon reasonable request.
### 8.5 Supplementary Document
Some theoretical equations are given in Supplement 1